\documentclass[11pt,preprint,tighten,flushrt]{aastex}

\newcommand{\beq}{\begin{equation}}
\newcommand{\eeq}{\end{equation}}
\newcommand{\beqa}{\begin{eqnarray}}
\newcommand{\eeqa}{\end{eqnarray}}

\def\IRAS{{\it IRAS\/} }

\def\eq#1{equation~(\ref{#1})}
\def\lexp{\mathop{{\langle}}\nolimits}
\def\rexp{\mathop{{\rangle}}\nolimits}
\def\dD{\delta_{\rm D}} 
\def\dt{\tilde \delta}
\def\Ft{\tilde F}
\def\d{\delta}

\font\BFd=cmmib10 scaled 1095
\font\BFt=cmmib10 scaled 1095
\font\BFs=cmmib10 scaled 800
\font\BFss=cmmib10 scaled 700

\def\bb#1{\relax
\ifmmode\mathchoice
{{\hbox{\BFd #1}}}
{{\hbox{\BFt #1}}}
{{\hbox{\BFs #1}}}
{{\hbox{\BFss #1}}}
\else\mbox{#1}\fi }

\def\k{{\bb{k}}}
\def\x{{\bb{x}}}
\def\z{{\bb{z}}}

\def\Jy{\,{\rm Jy}}
\def\Mpc{\,h^{-1}\,{\rm Mpc}}
\def\kMpc{\,h\,{\rm Mpc}^{-1}}

\font\ss=cmss10 scaled 1095
\def\P{\hbox{\ss P}}

\begin{document}

\title{The Bispectrum of $\bb{IRAS}$ Redshift Catalogs}

\author{Rom\'{a}n Scoccimarro,\altaffilmark{1,2} 
Hume A. Feldman,\altaffilmark{3} J. N. Fry,\altaffilmark{4} 
and Joshua A. Frieman\altaffilmark{5,6}}

\altaffiltext1{Institute for Advanced Study, School of Natural Sciences,
Einstein Drive, Princeton, NJ 08540; scoccima@ias.edu (current address)}

\altaffiltext2{CITA, McLennan Physical Labs, 60 St George Street,
Toronto ON M5S 3H8, Canada}

\altaffiltext3{Department of Physics \& Astronomy, University of Kansas, 
Lawrence KS 66045; feldman@ukans.edu}

\altaffiltext4{Department of Physics, University of Florida, 
Gainesville FL 32611-8440; fry@ufl.edu}

\altaffiltext5{NASA/Fermilab Astrophysics Center, Fermi National 
Accelerator Laboratory, Batavia, IL  60510; frieman@fnal.gov}

\altaffiltext6{Department of Astronomy and Astrophysics, University 
of Chicago, Chicago, IL 60637}

\begin{abstract}

We compute the bispectrum for the galaxy distribution in the \IRAS
QDOT, 2~Jy, and 1.2~Jy redshift catalogs for wavenumbers $ 0.05 \leq k
\leq 0.2 \kMpc $ and compare the results with predictions from
gravitational instability in perturbation theory.  Taking into account
redshift space distortions, nonlinear evolution, the survey selection
function, and discreteness and finite volume effects, all three
catalogs show evidence for the dependence of the bispectrum on
configuration shape predicted by gravitational instability.  Assuming
Gaussian initial conditions and local biasing parametrized by linear
and non-linear bias parameters $b_1$ and $b_2$, a likelihood analysis
yields $1/b_1 = 1.32^{+0.36}_{-0.58}$, $1.15^{+0.39}_{-0.39}$ and
$b_2/b_1^2=-0.57^{+0.45}_{-0.30}$, $-0.50^{+0.31}_{-0.51}$, for the
for the 2~Jy and 1.2~Jy samples, respectively. This implies that \IRAS
galaxies trace dark matter increasingly weakly as the density contrast
increases, consistent with their being under-represented in clusters.
In a model with $\chi^2$ non-Gaussian initial conditions, the
bispectrum displays an amplitude and scale dependence different than
that found in the Gaussian case; if \IRAS galaxies do not have bias $
b_{1}> 1$ at large scales, $\chi^2$ non-Gaussian initial conditions
are ruled out at the 95\% confidence level.  The \IRAS data do not
distinguish between Lagrangian or Eulerian local bias.

\end{abstract}

\subjectheadings{cosmology: observations --- 
methods: statistical --- large-scale structure of universe}

\mbox{}

\section{Introduction}

Observations suggest that the distribution of matter on cosmological
scales is nearly homogeneous, with structure that in our present
understanding has grown by gravitational instability from small,
nearly Gaussian, initial fluctuations.  On large scales, departures
from homogeneity are small, and their evolution can be studied in
perturbation theory.  To linear order in perturbation theory, an
initially Gaussian fluctuation distribution remains Gaussian, but at
higher orders this is no longer true.  Details of the departures
depend on the nonlinear couplings; thus, non-Gaussianity can probe the
dynamics of structure formation.  One useful non-Gaussian measure is
the three-point correlation function, or the bispectrum in the Fourier
domain, a quantity that vanishes identically for a Gaussian
distribution but is nonzero and has a characteristic behavior as a
function of the shape of the triangle defined by its three wavevector
arguments if large-scale structure is formed as a result of
gravitational instability \citep{F84}.  When applied to the galaxies,
the bispectrum also depends on galaxy biasing, i.e. on the relation
between the galaxy and mass distributions \citep{FG93,F94} and on any
non-Gaussianity of the primordial density fluctuations \citep{FS94}.

Early measurements of the bispectrum in the Lick galaxy catalog did
not find the expected gravitational instability signature
\citep{FS82}.  In angular catalogs such as Lick, the density field is
integrated over the radial selection function, which in principle can
change the dependence on shape, although it appears that in practice
the effects of projection on the shape dependence of the bispectrum
are not large \citep{FT99,BKJ00}.  Results on the moderate scales
probed by this catalog are also affected by nonlinear evolution beyond
leading order in perturbation theory, which weakens the configuration
dependence of the bispectrum in a way that mimics the effects of bias
\citep{SCFFHM98}.  A more recent determination of the angular
three-point correlation function in the APM survey \citep{FG99}, which
probes significantly larger scales than Lick, shows an amplitude and
configuration dependence consistent with gravitational instability
from Gaussian initial conditions and no bias, though with large
errors.

Recent galaxy redshift surveys have mapped the distribution of
galaxies out to large distances.  The largest distances are obtained
for deep pencil beam and slice surveys; however, surveys with highly
anisotropic one- and two-dimensional geometries have window functions
which extend to large wavenumber in one or two directions,
corresponding to the smallest survey dimension(s). As a consequence,
nonlinearity on this scale tends to wash out the configuration
dependence of three-point statistics.  Thus, although the three-point
correlation function has been measured in the Las Campanas Redshift
Survey \citep{JB98}, the survey geometry is effectively
two-dimensional, with redshift slices of thickness about $ 10 \Mpc $;
this scale is not large enough to allow a direct comparison with
perturbation theory.  Thus, it is very difficult to extract
quantitative information from this survey about biasing and the nature
of the initial conditions.

To compare with predictions of the configuration dependence of the
bispectrum from perturbation theory, we wish to observe the bispectrum
directly in an analysis that to the extent possible is not affected by
the complications of projection and nonlinear evolution.  In this
paper, we report measurement of the bispectrum in several galaxy
redshift catalogs derived from the \IRAS Point Source Catalog 
\citep{IRASPSC}.  Working
with three-dimensional (redshift) data avoids the effects of radial
projection.  Moreover, the large volume and nearly full sky coverage
of these surveys allow us to make measurements at small wavenumber
(large scales), where clustering is well-described by leading order
perturbation theory.  Although working in redshift space introduces
distortions arising from the modulation of position induced by
peculiar velocities, these can be computed 
\citep{HBCJ95,VHMM98,SCF99}.

In Section~2 we review the leading order nonlinear perturbation theory
predictions for the bispectrum, including its dependence on
configuration shape, the distortions introduced by peculiar velocities
in redshift space, and the effects of a bias in the galaxy
distribution relative to matter.  In Section~3 we summarize properties
of the {\it IRAS}-based redshift catalogs, and we present our
estimators in Fourier space, using the optimal weighting of
\citet[FKP]{FKP94}.  In Section~4 we summarize the data analysis
method based on comparison with mock catalogs developed in the
companion paper \citep[Paper I]{S00b}, and in Section~5 we present the
results of our analysis.  Section~6 contains a final discussion.

\section{The Bispectrum in Perturbation Theory}

We compute statistics in the Fourier domain, since it naturally 
separates clustering on different physical scales.
The Fourier amplitude $\dt(\k)$, the transform of the  
fractional density contrast $ \d(\x) $, is defined by 
\beq
\dt(\k) = \int d^3 x \, \d(\x) \, e^{-i\k \cdot \x} , \qquad 
\d(\x) = \int {d^3 k \over (2\pi)^3} \, \dt(\k) \, e^{i\k \cdot \x} , 
\eeq
where the mass density is $ \rho(\x) = \bar \rho \, [1 + \d(\x)] $.
The first moments of $ \dt(\k) $ are the power spectrum 
and the bispectrum,
\beqa
\lexp \dt(\k_1) \dt(\k_2) \rexp &=& 
(2 \pi)^3 \dD(\k_1+\k_2)\, P(k_1) , \\
\lexp \dt(\k_1) \dt(\k_2)  \dt(\k_3) \rexp &=& 
(2 \pi)^3 \dD(\k_1+\k_2+\k_3) \, B(\k_1,\k_2,\k_3) .
\eeqa
The ``momentum-conserving'' Dirac $\delta$-function reflects 
statistically the underlying homogeneity and implies that the 
bispectrum is defined for configurations of wavenumbers that form 
closed triangles. If the initial fluctuations are Gaussian, 
then the bispectrum vanishes in linear perturbation theory; 
however, mode-mode couplings when nonlinearities 
are included generate a non-zero $B$ even for an initially 
Gaussian distribution.

In gravitational instability with Gaussian initial conditions, 
the bispectrum at second (leading nonzero) order in 
perturbation theory is \citep{F84} 
\beq 
B_{123} = Q_{12} P_1 P_2 + Q_{13} P_1 P_3 + Q_{23} P_2 P_3 ,
\label{B123}
\eeq
where $ P_1 $ denotes $ P(k_1)$, etc., 
and $B_{123} = B(k_1,k_2,k_3) $. Here 
$ Q_{ij} $ is determined entirely by the configuration geometry, 
\beq
Q_{ij} = (1 + \kappa) + \cos \theta_{ij} 
\left( {k_i \over k_j} + {k_j \over k_i} \right) 
+ (1 - \kappa) \cos^2 \theta_{ij} , 
\label{Qij}
\eeq
where $ \cos \theta_{ij} = \hat \k_i \cdot \hat \k_j $.  The factor
$\kappa$ is $ \kappa = {3 \over 7} $ in an Einstein-de Sitter
universe, with critical matter density $\Omega_m=1$, and it never
varies much from this value.  For an open universe with $ \Omega_m $
near 1, $ \kappa \approx {3 \over 7} \Omega_m^{-2/63} $, and for a
flat universe with cosmological constant, $ \kappa \approx {3 \over 7}
\Omega_m^{-1/143} $ \citep{BJCP92,BCHJ95}; while in both cases $
\kappa \to \frac12 $ as $ \Omega_m \to 0 $.

It is useful to define the reduced bispectrum, 
\beq
\label{Qdef}
Q \equiv { B_{123} \over P_1 P_2 + P_1 P_3 + P_2 P_3 } . 
\eeq
For a power-law spectrum, $P \propto k^n$, at second order in
perturbation theory, the reduced bispectrum $Q$ is independent of time
and of scale leaving only the dependence on triangle shape.  This
behavior contrasts sharply with that of $P$ and $B$, which vary
strongly with scale. Moreover, the expressions above imply that $Q$ is
very insensitive to the value of $\Omega_m$.  The solid curve in
Figure~\ref{Qpt} shows the characteristic dependence of $Q$ on
configuration shape, here parameterized by the angle $\theta_{12}$
between two sides of ratio $ k_2/k_1 = {1 \over 2}$, for a power-law 
spectrum with spectral index $ n = -1.4 $ and for $ \Omega_m = 1 $.  
This ratio
of sides minimizes the total range of scales covered as $ \theta $
varies from 0 to $\pi$, but in fact $Q$ is insensitive to the ratio $
k_2/k_1 $ for $ \theta < \pi/2 $.
The characteristic dependence on shape, with an enhancement 
for colinear configurations ($ \theta = 0 $ or $ \theta = \pi $), 
reflects the expected anisotropy of gravitational collapse.
The dotted curve shows the behavior for $ \Omega_m \to 0 $.  Other
curves show redshift space and bias results, which we discuss next.

In redshift space, where radial distances are assigned from recession 
velocities by Hubble's law, peculiar velocities induce distortions in 
clustering statistics (Kaiser 1987; for a review, see Hamilton 1997).  
In the plane parallel approximation, where the line of sight is taken 
as a fixed direction $\hat \z$, the density in space $\dt(\k)$ and 
redshift space $\dt_{s}(\k)$ are related to linear order by 
$ \dt_s  (\k) = [ 1 + f_1 (\Omega) \mu^{2} ]\ \dt(\k) $, 
where $ f_1 = \Omega^{4/7} $ and $ \mu \equiv k_{z}/k $.
Averaging over orientations, the power spectrum monopole 
in redshift space then becomes 
\beq  
P_s (k) = \left( 1 + {2 \over 3} f_1 + {1 \over 5} f_1^2 \right) \, P(k) , 
\eeq 
while the bispectrum monopole is of the same form as as in 
\eq{B123} above, with \citep{HBCJ95} 
\beqa Q_{ij,s} &=& a_0 + a_2 \cos^2 \theta_{ij} + a_4 \cos^4 
\theta_{ij} + (a_1 \cos \theta_{ij} + a_3 \cos^3 \theta_{ij}) \left( 
{k_i \over k_j} + {k_j \over k_i} \right) \nonumber \\
&&\qquad {} + (1 - \cos^2 \theta_{ij})
\left[ {x_1 (k_i^2 + k_j^2) + x_2 k_i k_j \cos \theta_{ij}} \over
{ k_i^2 + k_j^2 + 2 k_i k_j \cos \theta_{ij} } \right] \label{Qred}
\eeqa
where the coefficients are 
\beqa
\textstyle
a_0 &=& (1+\kappa) + \frac23 (2+\kappa) f_1 + \frac1{15}(11+\kappa)
f_1^2
+ \frac6{35}f_1^3 + \frac6{315}f_1^4
+ \kappa f_2 \left( \frac13 + \frac6{15} f_1 + \frac1{35} f_1^2 \right)
\\
a_1 &=& 1 + \frac43 f_1 + \frac{14}{15}f_1^2 + \frac{12}{35}f_1^3
+ \frac{15}{315} f_1^4 \\
a_2 &=& (1-\kappa) + \frac23 (2-\kappa) f_1 + \frac1{15}(23+\kappa)
f_1^2
+ \frac{30}{35}f_1^3 + \frac{48}{315}f_1^4
- \kappa f_2 \left( \frac13 + \frac6{15}f_1 - \frac3{35}f_1^2 \right) \\
a_3 &=& \frac4{15} f_1^2 + \frac8{35} f_1^3 + \frac{20}{315} f_1^4 \\
a_4 &=& \frac2{15} (1-\kappa)f_1^2 + \frac4{35}f_1^3
+ \frac{16}{315} f_1^4 - \frac4{35} \kappa f_1^2 f_2 \\
x_1 &=& - \frac2{15} \kappa f_1 f_2 , \qquad 
x_2 =  \frac4{35} \kappa f_1^2 f_2 ; 
\eeqa
$ \kappa $ is as given before, and $f_1$ and $f_2$ are logarithmic 
derivatives of the first and second order perturbation theory growth 
factors $ D_1 $ and $ D_2 $ with respect to the cosmological scale 
factor, $ f_1 = d \log D_1 / d \log a $, etc.  For $ \Omega_m $ near 1, 
$ f_1 \approx \Omega^{4/7} $, and $ f_2 \approx 2 f_1 $.  Setting $ 
f_1 = f_2 = 0 $ gives $ a_0 = 1 + \kappa $, $ a_1 = 1 $, $ a_2 = 1 - 
\kappa $, and $ a_3 = a_4 = x_1 = x_2 = 0 $, reproducing \eq{Qij}, 
the result in real space.

Figure~\ref{Qpt} shows the predictions for $ P \sim k^n $ with $ n =
-1.4 $, a good approximation for the \IRAS samples over 
the range of scales we use.
The long-dashed curve
shows the perturbation theory result for $ Q_s(\theta) $ in redshift
space for $ \Omega_m=1 $; the short-dashed curve $ \Omega_m = 0.3 $.
The behavior depends on $ \Omega_m $, but more weakly than the factor
$ f(\Omega) = \Omega^{4/7} $ that appears in all peculiar velocities.
Between $ \Omega = 1 $ and $ \Omega =
0.3 $, $ f $ varies by a factor of 2, but $Q$ by at most about 15\%, 
while 
the redshift space result for $ \Omega_m \to 0 $ is the same as in
real space.  Over the entire range of $ \Omega $ the characteristic
behavior is very much the same.

To translate these results for the mass distribution into predictions 
for the galaxy distribution, we must take into account the effects of 
bias. In the simplest case, a local 
deterministic bias model, in which the galaxy density is a function of 
the local mass density, the galaxy number contrast can be expressed as 
a series in the mass density contrast,
\beq
\delta_g(\x) = \sum_{k=0}^\infty {b_k \over k!} \delta^k(\x) .
\label{locbias}
\eeq
To leading order in perturbation theory, the resulting galaxy power 
spectrum  is simply related to the mass power spectrum, 
$ P_g (k)= b_1^2 \, P(k) $, where $b_1 $ is the linear bias factor.  
In this model, the galaxy bispectrum amplitude is  
\citep{FG93,F94} 
\beq
Q_g = {Q \over b_1}  + {b_2 \over b_1^2} . \label{Qbias}
\eeq
Thus, the effect of $ b_{1} > 1 $ is to reduce the 
dependence of $Q$ on triangle shape, while $ b_2 $ introduces 
a constant offset.
Since biasing operates in real space, instead of redshift space, 
\eq{Qbias} does not hold for redshift-space quantities; however, it 
turns out that the deviations from \eq{Qbias} are small \citep{SCF99}.
The dot-dashed curve in Figure~\ref{Qpt} shows the effect on $Q$ 
expected for $ 1/b_1 = 1.25 $ and $ b_2/b_1^2 = -0.5 $.

We also compare our results below with previous determinations of the
skewness $ S_3 $.  One-point moments of the galaxy overdensity in a
cell of volume $V$, $\bar \xi_p = \lexp \delta_g^p \rexp_c $,
correspond to integrals over the $p-$point correlation functions
$\xi_{p}$,
\beq
\bar \xi_p = {1 \over V^p} \int_V d^3x_1 \ldots d^3x_p \, \xi_p .
\eeq
These moments also display hierarchical scaling to leading 
order in nonlinear perturbation theory, 
\beq
\bar \xi_p = S_p \, \bar \xi^{p-1} , 
\eeq
where the parameters $S_p$ depend on the shape of the volume 
and on the spectral index.
For a spherical tophat filter and power-law spectrum 
$ P \sim k^n $, the skewness $ S_3 $ is 
\citep{FG93} 
\beq
S_3 = {1 \over b_{1}} \left[ {34 \over 7} - (3+n) \right] 
+ {3b_2 \over b_{1}^2} ,
\label{S3}
\eeq
where the result in brackets corresponds to the skewness in the mass  
induced by gravity from Gaussian initial conditions \citep{JBC93,B94}.

The relation of \eq{Qbias} assumes the bias is local and
deterministic.  Locality here means that the scales we consider below,
$ k \leq 0.2 \kMpc $, are assumed to be much larger than the
characteristic scale of galaxy formation, where nonlocal processes
other than gravity become important.  Deterministic bias assumes that
the scatter in the relation between $\delta_g$ and $\delta$ can be
neglected, with biasing described by the {\em mean} relation in
\eq{locbias}.  Deviations from deterministic bias can be introduced by
adding a random field $\epsilon$, whose variance describes the scatter
about the mean, to the right side of \eq{locbias}.  However, at least
for models in which the scatter is local, that is, uncorrelated at
large scales, $\lexp \epsilon (\x_{1}) \epsilon (\x_{2}) \rexp \propto
\dD(\x_{1}-\x_{2})$, the use of \eq{locbias} together with bispectrum
measurements to constrain bias parameters results in a good recovery
of the mean biasing relation (Paper~I).

We note that the use of the bispectrum to recover information about
galaxy bias does not depend exclusively on the validity of \eq{Qbias};
other biasing schemes may predict other relationships between galaxy
and dark matter bispectra that can similarly be tested. For example,
in non-Gaussian models, bias and non-Gaussianity interact in a
non-trivial way \citep{FS94}.  We will consider an example of this
class of models below when we discuss $\chi^{2}$ initial conditions.
Alternatively, one can consider models in which the bias is local in
Lagrangian space (e.g. \cite{MoWh96}) rather than Eulerian space as
assumed in \eq{locbias}.  In this case, the conditions for galaxy
formation depend on the linearly extrapolated density field [and thus
\eq{locbias} holds for the initial field configurations which are then
evolved by gravity], and the prediction for the reduced bispectrum in
\eq{Qbias} changes to \citep{CLMP98} 
\beq Q_g = {1 \over b_1} Q_{123}
+ {b_2 \over b_1^2} + \left(\frac{b_{1}-1}{b_{1}^{2}}\right) {\Delta
Q_{12} P_{1}P_{2} + \Delta Q_{23} P_{2}P_{3} + \Delta Q_{31}
P_{3}P_{1} \over P_{1}P_{2} + P_{2}P_{3} +
P_{3}P_{1}}\label{Qbiaslag}, 
\eeq 
where $\Delta Q_{ij} = (k_{i}/k_{j}+k_{j}/k_{i}) \cos\theta_{ij}$.
The additional source of shape dependence in the third term can then
be tested against observations. It is interesting to note that the
difference between Eulerian and Lagrangian bias does not show up in
large-scale measurements of the power spectrum; to leading (linear)
order, both schemes agree. It is the sensitivity of the bispectrum to
non-linear effects that makes possible to distinguish between these
schemes.

\section{Catalogs and Optimally Weighted Fourier Transform}

We apply our analysis to three \IRAS redshift subsamples: the updated
QDOT survey \citep{EKSLREF90} and the 2~Jy \citep{SHDYFT92} and 1.2~Jy
\citep{FHSDYS95} redshift surveys.  The QDOT survey chooses at random
one in six galaxies from the \IRAS point source catalog with a $ 60 \,
\mu{\rm m} $ flux $ f_{60} > 0.6 \Jy $.  In this sample there are 1824
galaxies with galactic latitude $ |b| > 10 \arcdeg $, with redshifts
that correspond to distances $ 20 \Mpc < R < 500 \Mpc $.  We have used
a revised version of the QDOT database in which a redshift error that
afflicted approximately 200 southern galaxies has been corrected, and
we have converted all redshifts to the local group frame.  The other
two samples are shallower but denser than the QDOT catalog.  The 2~Jy
catalog, complete to a flux limit $ f_{60} > 2 \Jy $, contains 2072
galaxies; the 1.2~Jy catalog, with $ f_{60} > 1.2 \Jy $, contains 4545
galaxies.  Although all three catalogs are large-angle surveys, they
are not full-sky: because of obscuration due to the Galactic plane,
the existence of some bright sources, and other technical details,
these surveys cover a solid angle of $ 9.282 \, {\rm sr} $, about
three quarters of the sky. In a subsequent paper (Paper III) we will
analyze the IRAS PSCz survey \citep{PSCz}.

To deal with the angular mask and the redshift distribution, we follow
the formalism developed by FKP, computing moments of a weighted
transform of the difference between the data and a synthetic random
Poisson catalog with the same radial distribution and angular mask,
\beq
\Ft(\k) = \int d^3x \, w(\x) \, 
[ n_g(\x) - \alpha n_s(\x) ] \, e^{-i \k \cdot \x} , 
\eeq
where $ n_g(\x) = \sum_i \dD (\x - \x_i) $ is the point distribution
for the galaxy data and $ n_s(\x) $ similarly for the synthetic
catalog.  The factor $ \alpha = N_g / N_s $ scales the synthetic
catalog to the same density as the data, and $ w(\x) = 1 / [ 1 + {\bar
n}(\x) P_0 ] $ is the ``optimal'' weight function for computing the
power spectrum (and bispectrum as well, see Paper I); for detailed
analysis of the formalism we follow, see FKP.  The subtraction removes
a contribution to the power from the shape of the radial selection
function.  The Fourier transform is performed precisely, summed over
objects at their exact positions, rather than using interpolation and
a fast Fourier transform, for frequencies that are uniformly spaced on
a grid, $ \k = (n_x, n_y, n_z) \, k_f $, where $ k_f = 0.005 \kMpc $.
The results in this paper are obtained by setting the fiducial power
in the weight function to $ P_0=8000 $, 2000, and $ 2000 \, (h^{-1}
{\rm Mpc})^3 $ for QDOT, 2~Jy, and 1.2~Jy respectively. These values
of $P_0$ reflect the different effective depths of the samples; in
practice, the inferred value of $Q$ is not very sensitive to this
parameter.

Including contributions from discreteness, the second moment of 
$ \Ft $ is related to the power spectrum of the galaxy distribution as 
\beq
\lexp |\Ft(\k)|^2 \rexp = \int {d^3k' \over (2\pi)^3} \, 
|G(\k')|^2 \, P(|\k-\k'|) + (1 + \alpha)  \int d^3x \, w^2 \bar n , 
\label{Fk1}
\eeq
where the window function $ G(\k) $ is 
\beq
G(\k) = \int d^3x \, w(\x) \bar n(\x)  \, e^{-i \k \cdot \x} ~, 
\eeq
and $ \bar n(\x) = \lexp n_g({\x}) \rexp $ is the mean space density
of galaxies, given the angular and luminosity selection criteria.  As
noted above, for anisotropic survey geometries the window function
$G(\k)$ is broad in at least one dimension; as a result, a range of
scales contribute to $ \lexp |\Ft(\k)|^2 \rexp$. For the \IRAS
catalogs, however, $ G(\k) $ is a rapidly falling function of $|\k|$;
in this case, for $k$ not too small, \eq{Fk1} becomes 
\beq \lexp
|\Ft(\k)|^2 \rexp = P(k) \int d^3x \, w^2 \bar n^2 + (1 + \alpha) \int
d^3x \, w^2 \bar n ~.  
\eeq 
Similarly, the three-point function of $\Ft$ becomes
\beq 
\lexp \Ft_1 \Ft_2 \Ft_3 \rexp = B_{123} \int d^3x
\, w^3 \bar n^3 + (P_1 + P_2 + P_3) \int d^3x \, w^3 \bar n^2 + (1 -
\alpha^2) \int d^3x \, w^3 \bar n .  
\eeq

Subtracting the discreteness contributions, our estimators for the
power spectrum and bispectrum are
\beq
\hat P(k) = 
{\lexp |\Ft(\k)|^2 \rexp \over \int d^3x \, w^2 \bar n^2} 
- (1 + \alpha) {\int d^3x \, w^2 \bar n \over \int d^3x \, w^2 \bar n^2} , 
\eeq
\beq
\hat B_{123} = 
{\lexp \Ft_1 \Ft_2 \Ft_3 \rexp \over \int d^3x \, w^3 \bar n^3} 
- (\hat P_1 + \hat P_2 + \hat P_3) {\int d^3x \, w^3 \bar n^2 
\over \int d^3x \, w^3 \bar n^3} - (1 - \alpha^2) 
{\int d^3x \, w^3 \bar n \over \int d^3x \, w^3 \bar n^3} , 
\eeq
and the reduced bispectrum estimator is 
\beq
\hat Q = { \hat B_{123} \over \hat P_1 \hat P_2 
+ \hat P_1 \hat P_3 +  \hat P_2 \hat P_3} . \label{Qestim}
\eeq


\section{Data Analysis: Mock Catalogs}

Having obtained the bispectrum amplitude $Q$ following the steps in
the previous section, we are faced with interpreting the results and
comparing to theoretical predictions such as those in Section~2.  To
do this, we must take into account a number of effects present in real
redshift surveys that can potentially modify these predictions.  In
this section we summarize results from Paper~I, where these effects
are discussed in detail using mock catalogs constructed from
perturbation theory and $N$-body simulations.  We refer the reader to
Paper~I for a more in-depth analysis.

\subsection{Nonperturbative Redshift Distortions}

There are two ways in which the perturbation theory calculations of
redshift space distortions described in Section~2 potentially need 
to be improved, namely, by going beyond both the plane-parallel and
perturbative redshift mapping approximations.  In the plane-parallel
approximation assumed above, the structures being observed are
considered to be sufficiently far away that the line of sight can be
considered a fixed direction.  Since the \IRAS surveys cover most of
the sky, this is not necessarily a good approximation.  In this paper,
we only consider the angle-averaged (monopole) power spectrum and
bispectrum, which are very insensitive to the radial nature of the
distortions (see Figs.~2 and 3 in Paper I).  In our likelihood
analysis, we nevertheless use mock catalogs in which the
redshift-space mapping is correctly performed radially.

By working at large scales, $ k \leq 0.2 \kMpc $, we can apply leading
order nonlinear perturbation theory to describe the shape depence of
the bispectrum in the absence of redshift distortions
\citep{SCFFHM98}.  However, in redshift space, another effect beyond
nonlinear dynamics enters: the mapping from real to redshift space is
itself is nonlinear.  The discussion of redshift distortions in
Section~2 assumed a perturbative expansion for this mapping as well as
the dynamics.  The validity of this expansion is in fact rather
restricted \citep{SCF99}; over the range of scales we consider,
nonlinear corrections to the redshift-space mapping are non-negligible
(see Fig.~3 in Paper I).  We take these effects into account by using
second order Lagrangian perturbation theory (2LPT) numerical
realizations in which the redshift-space mapping is done exactly.

\subsection{Finite Volume Effects}

The finite volume of a survey can alter the amplitude and scale or
shape dependence of the bispectrum, potential systematic effects that
we must consider before we can directly compare theory with
observations.
[For a general discussion of the same issues in one-point statistics, 
see \citet{SCB99} and references therein.] 
Since we work
at scales small compared to the extent of the survey, the integral
constraint bias is negligible, thus there are basically two
manifestations of finite volume effects: estimator bias and
statistical bias.

Estimator bias, the fact that the estimator of the reduced bispectrum
in \eq{Qestim} does not obey $ \lexp \hat Q \rexp = Q$, has two
significant sources.  Although the power spectrum and bispectrum
estimators are themselves unbiased, $\hat Q$ is a non-linear
combination of them, thus an estimation bias arises
\citep{HG99,SCB99}.  In Paper I, we found that taking into account the
estimator bias is important to meaningfully compare the bispectrum
obtained in different surveys.  However, this is not the full
story. One can consider these estimators as random fields, each of
which can be described by a probability distribution function (PDF)
and its moments. Normally, one requires an estimator to be unbiased (a
condition on its first moment) and of minimum variance (a requirement
on the second moment).  However, the higher order moments of
estimators can also play a role. For example, if the skewness of the
estimator PDF is positive (negative), even if the estimator is
unbiased, in a given realization the most likely value of the
estimator will be systematically lower (higher) than the mean.  We
call this {\em statistical} bias.  Since in observations we deal with
a single sample, this effect must be taken into account \citep{SC96}.  

We must also account for correlations between estimators.
In an infinite universe, Fourier modes are statistically independent.
For a survey of finite volume, translation invariance is broken,
leading to correlations between different Fourier modes at large
scales.  We confine our analysis to sufficiently small length scales,
$ k \ge 0.05 \kMpc $, that the window function of the
survey can be approximated by a delta function (see Paper I for a
detailed discussion).  However, Fourier modes will be correlated with
their neighbors in $k$-space over the inverse width of the selection
function of the survey.  In addition, correlations arising from shot
noise dominate at small scales.

A proper analysis of these surveys reqires we take these effects into
consideration.  In Paper~I we have studied this problem by using mock
catalogs drawn from numerical realizations of perturbation theory as
well as $N$-body simulations.  In order to measure the correlation
matrix of the bispectrum and the PDF of its estimator, many ($\geq
400$) realizations of the survey under consideration are necessary.
This is efficiently done using a numerical implementation of second
order Lagrangian perturbation theory (2LPT), which is orders of
magnitude faster than $N$-body simulations and reproduces the correct
bispectrum at large scales, including nonlinear aspects of redshift
distortions mentioned above.  In order to investigate the domain of
validity of 2LPT, we have run N-body simulations which confirm that,
over the range of scales we consider ($0.05 \leq k \leq 0.2 \kMpc $),
2LPT is a very good approximation to the full nonlinear theory.

\subsection{Likelihood Analysis}

The survey geometry and radial selection function introduce biases and
correlations that are best dealt with in a likelihood analysis.  We
proceed in the following way: Given a set of measured reduced
bispectrum amplitudes $ \{Q_{m}\} $, $m=1,\ldots,N_{T}$, where $N_{T}$
is the number of closed triangles in the survey, we diagonalize their
covariance matrix so that the $Q$-eigenmodes $\hat q_{n}$,
\beq
\hat{q}_n = \sum_{m=1}^{N_T} \gamma_{mn}\ \frac{Q_m-\bar{Q}_m}{\Delta 
Q_m},  
\eeq
satisfy
\beq 
\lexp \hat{q}_n \hat{q}_m \rexp = \lambda_n^2 \, \d_{nm}, 
\eeq
where $\bar{Q} \equiv \lexp Q \rexp$ and $ (\Delta Q)^2\equiv
\lexp Q^2 \rexp -\bar{Q}^2$.
These $Q$-eigenmodes have ``signal to noise'' ratio 
\beq 
\left( \frac{S}{N} \right)_n \equiv \frac{1}{\lambda_n} \, \left| 
\sum_{m=1}^{N_T} \gamma_{mn} \, \frac{\bar{Q}_m}{\Delta Q_m} \right| .
\eeq
The physical interpretation of the $Q$-eigenmodes becomes clear when
ordered in terms of their signal to noise.  The ($n=1$) eigenmode with
highest signal to noise, $S/N \approx 3$, corresponds to all weights
$\gamma_{m1}>0$; it represents the overall amplitude of the
bispectrum.  The $n=2$ $Q$-eigenmode, with $S/N \approx 1$, has
$\gamma_{m2}>0 $ for nearly flat triangles and $\gamma_{m2}<0$ for
nearly equilateral triangles---it represents the configuration
dependence of the bispectrum.  Higher order ($n>2$) eigenmodes contain
further information, such as variations of the amplitude and shape
with scale; although they generally have $S/N < 1$, there are many of
them, so they cannot be neglected.

If the joint PDF of the $Q_{m}$ were Gaussian, then diagonalizing the
covariance matrix would guarantee that the new eigenmodes are
statistically independent---the full PDF would be the product of their
individual PDF's. Although this is not necessarily so in the general
non-Gaussian case, we will assume that this independence is a good
approximation here (supported by the results of Paper~I).  We can then
write down the likelihood as a function of the parameters, 
\beq 
{\cal L}(\alpha_{1},\alpha_{2}) \propto \prod_{i=1}^{N_T}
\P_{i}[\nu_i(\alpha_{1},\alpha_{2})],
\label{like}
\eeq
where $\alpha_{1}\equiv 1/b_1$, $\alpha_{2} \equiv b_2/b_1^2$, and
\beq 
\nu_i(\alpha_{1},\alpha_{2}) \equiv \frac{1}{\lambda_{i}}\ 
\sum_{j=1}^{N_T} \gamma_{ji} \, \frac{Q_j - (\alpha_{1} 
\bar{Q}_j+\alpha_{2})}{\Delta Q_j} ~.
\label{nu}
\eeq Here the $Q_j$ ($j=1,\ldots,N_{T}$) are the data, and the mean
$\bar{Q}_j$, standard deviation $\Delta Q_j$, and the non-Gaussian
PDF's $\P_{i}(\nu_{i})$ are extracted from the mock catalogs.  As
described in Section~2, the reduced bispectrum is mostly sensitive to
the power spectrum shape, with a very small dependence on $\Omega_m$
through redshift distortions and negligible dependence on the
normalization $\sigma_{8}$ at large scales.  Our mock catalogs are
based on the $\Lambda$CDM model, with $\Omega_m=0.3$,
$\Omega_{\Lambda}=0.7$, and $\sigma_{8}=0.7$; in redshift space, for
this model $\sigma_{8}=0.84$, in agreement with the observed
normalization for \IRAS galaxies \citep{FDSYH94}.  The power spectrum
shape of this model is also in good agreement with the observed \IRAS
galaxy power spectrum over the range of scales we consider
\citep{FKP94,FDSYH93}.  If we change the power spectrum shape, the
resulting bias parameters do change, as the prediction for $Q$ is
sensitive to spectral index; for example, for a shape parameter
$\Gamma=0.5$, as in the SCDM model, $Q$ is related to that in the
$\Lambda$CDM model ($\Gamma=0.21$) by approximately $Q(\Gamma=0.21)
\approx 1.3 Q(\Gamma=0.5)-0.05$ (Fig.~5, Paper I), so that $b_1^{
SCDM} \approx b_1^{\Lambda CDM}/1.3$ and $(b_2/b_1^2)^{SCDM} \approx
(b_2/b_1^2)^{\Lambda CDM} -0.05/b_1^{SCDM}$; thus, the inferred 
bias parameters for \IRAS galaxies would be smaller if we assumed SCDM
rather than the $\Lambda$CDM model.  To assess the sensitivity of our
results to a change in $\Omega_m$, we also consider mock catalogs for
a model with $\Omega_m=1$ but with the same power spectrum
normalization and shape as the $\Lambda$CDM model above.  We do not
consider mock catalogs with different values of $\sigma_{8}$, since
the dependence of $Q$ on this parameter is very weak (see Paper~I).
The main effect of a change in $\sigma_{8}$ would be to decrease
(increase) the error bars we infer if \IRAS galaxies have linear bias
parameter $b_{1}<1$ ($b_{1}>1$), due to shot noise reduction
(increase).  In this regard, the error bars we derive on the bias
parameters can be considered conservative.

Important results obtained from results of mock catalogs as 
described in Paper~I include 
\begin{itemize}
\item Estimator bias can be as large as a factor of two.  However,
\eq{Qbias} continues to hold for biased estimates.
\item The PDF of the bispectrum estimator is generally skewed and has 
non-Gaussian exponential tails or even power-law tails in the case 
of sparse sampling (QDOT) and $\chi^{2}$ initial conditions.
\item The cross-correlation coefficient induced by the window function
of the survey and shot-noise between different power spectrum bins or
different triangles is not significantly less than unity.
\end{itemize}

A maximum likelihood method to constrain bias parameters 
from redshift surveys has been proposed before by \cite{MVH97}.
However, their method does not take into account these
effects and thus cannot be applied as it stands to \IRAS surveys.

\section{Results}

The bispectrum is defined for closed triangles in Fourier space.
After averaging over orientations, we can characterize a given
triangle by the lengths of its three sides, $ k_1 $, $ k_2 $, $ k_3 $,
where we order $ k_1 \ge k_2 \ge k_3 $, or by any combination of
parameters that completely defines a triangle.  Figures~\ref{QkQDOT},
\ref{Qk2Jy}, and \ref{Qk1.2Jy} show the scale dependence of $ Q(k_1) $
for the QDOT, 2~Jy, and 1.2~Jy surveys, for triangles in which the two
shorter sides $ k_2 $ and $ k_3 $ are separated by angle $
\theta_{23}$, i.e., $ \hat \k_2 \cdot \hat \k_3 = \cos \theta_{23} $.
The four panels show results for four ranges of the angle
$\theta_{23}$; small $ \theta_{23} $ corresponds to nearly flat
triangles, i.e., all three sides almost colinear, while triangles with
larger $\theta_{23}$ are more open.  The results at very small $k$ are
sensitive to the subtraction of power in the selection function, so we
have constrained the smallest side of the triangle to have $ k_3 \geq
0.05 \kMpc $.  The scatter in the values of $Q$ is not small, but it
is still apparent that there is a nonzero positive signal.
Neighboring points are clearly correlated; for this reason we have not
included error bars that describe the dispersion of values that enter
into each point.  A full treatment of triangle correlations, as
described in \S~4, is carried out in the analysis below.  Dashed lines
show the median and 68\% width for the points in each panel.  The
average value of $Q$ in each window generally decreases as
$\theta_{23}$ increases, as expected from gravitational instability, 
even for QDOT, where the scatter is largest.

Figures \ref{QaQDOT}, \ref{Qa2Jy}, and \ref{Qa1.2Jy} show the shape
dependence $ Q(\theta) $ for triangles with sides with two sides of
ratio 0.4--0.6 separated by angle $ \theta $.  The solid curves show
the tree-level (leading order) prediction from gravitational
instability in perturbation theory with a spectral index $n=-1.4$ 
(as in FKP), including perturbative redshift distortions 
(eq.~[\ref{Qred}]).
The long-dashed curves show $Q$ averaged over a large number of 
synthetic 2LPT catalogs, and includes nonperturbative effects.
Different symbols show results for $ k_1 = 0.1 $--$0.125
\kMpc $ (triangles), $ k_1 = 0.125 $--$ 0.15 \kMpc $ (squares), $ k_1
= 0.15 $--$ 0.175 \kMpc $ (diamonds), $ k_1 = 0.175 $--$ 0.2 \kMpc $
(circles).  The signal is weak in the sparse and noisy QDOT survey, 
but in both the 2~Jy and 1.2~Jy surveys a shape dependence is apparent, 
although there is significant dispersion, especially at 
nearly colinear configurations.
For 2~Jy and 1.2~Jy, open symbols show the direct result, and 
filled symbols are corrected for the average finite volume bias.
The short-dashed curves show the local bias  model of 
\eq{Qbias} using the results of the likelihood analysis below, 
including non-linear redshift distortions, estimator bias,
correlations between triangles, and the non-Gaussianity of the
bispectrum PDF, applied to the 2LPT result.
While the finite sample correction removes the bias in the mean, 
the distribution is still skewed to low values, and so the filled 
symbols cannot be directly compared to theoretical predictions.
It is visually apparent that the skewness is larger in the 
2~Jy catalog, consistent with the findings in Paper I.

To assess to what extent the data follow the shape dependence
predicted by gravitational instability and to extract quantitative
bias parameters, we have carried out the likelihood procedure
described in Section~4.  There are potentially 4741 closed triangles
binned with integer sides $k_i/k_f=10,\ldots,40$ (although the galaxy
overdensity is not mapped into a grid, we use wavevectors integer
multiples of $k_f=0.005\kMpc$).  Since the window function of the
survey has a width of order $\Delta k/k_f \approx 5$, the number of
``independent'' triangles is much smaller than 4741.  We therefore use
coarser bins, defined in terms of shape $s$, ratio $r$, and scale $k$
parameters \citep{P80}, \beq s \equiv \frac{k_1-k_2}{k_3}, \qquad r
\equiv \frac{k_2}{k_3}, \qquad k \equiv k_3,
\label{Qvar}
\eeq
where the ``shape'' parameter $s$ obeys $0\leq s \leq 1$ 
($k_1\geq k_2 \geq k_3$), the ratio $ 1 \leq r \leq 4$,  
and  the overall scale of the triangle satisfies $10 \leq k/k_f \leq 40$.
The shape parameter is $s=0$ for isosceles triangles 
(and if $r=1$ these are equilateral triangles); 
for $s=1$ we have colinear triangles.
Using 10 bins in each variable plus the closed triangle constraint 
yields 203 triangles.
We find that using a finer binning leads to a more unstable behavior of
numerical computation of the $Q$ eigenvalues and eigenvectors, while
making the binning too coarse loses information on the
shape and scale dependence of the bispectrum.

Figure~\ref{likeG} shows the results of the likelihood analysis 
for the 2~Jy and 1.2~Jy catalogs.
Due to the sparse sampling (and thus large shot noise errors), 
the bias parameters cannot be significantly constrained from the
QDOT survey, so we do not show results for it. The bottom panel in
Figure~\ref{likeG} shows 68\% and 90\% likelihood contours as a function
of $1/b_1$ and $b_2/b_1^2$, assuming Gaussian initial conditions and
matter density $\Omega_m=0.3$. The maximum likelihood analysis yields 
the results in Table~1, 
 with error bars obtained from the 68\% marginalized likelihoods shown
in the upper panels of Figure~\ref{likeG}.  Despite the apparently
different amplitudes of $Q$ in Figures~\ref{Qa2Jy} and \ref{Qa1.2Jy},
the bias parameters obtained for the two catalogs are remarkably
similar. The differences between Figures~\ref{Qa2Jy} and \ref{Qa1.2Jy}
are largely due to different finite volume effects in the two surveys,
which trace to their different selection functions. In particular, the
shallower 2~Jy catalog is more strongly affected by finite volume
corrections and skewness of the bispectrum PDF than the 1.2~Jy survey.

Figure~\ref{likeGo} shows the effect of changing the matter density
from $\Omega_m=0.3$ to $\Omega_m=1$ in the case of the 1.2Jy catalog
(similar results hold for the 2~Jy catalog). As expected from the
analytic results described in Section~3 and illustrated in
Figure~\ref{Qpt}, the results given in Table~1 change by about $10 \%$ 
 in the direction consistent with Figure~\ref{Qpt}: 
for $\Omega_m=1$ the expected configuration dependence
is slightly stronger, and thus the bias must be slightly larger to
account for the same measured value. Note however that the $\Omega_m$
dependence is not nearly as strong as in redshift-space power spectrum
measurements which determine the combination $\beta \approx
\Omega^{0.6}/b_1$; to retain the same value of $\beta$, our result
would have to come down to $1/b_1=0.56$ for $\Omega_m=1$. Thus, to a
large extent, measurements of the bispectrum can be used to lift the
degeneracy present in the power spectrum.

We found it useful to track how the results change as we vary 
the number of eigenmodes kept in the analysis. 
For the two highest $S/N$ eigenmodes,
which represent the overall amplitude and configuration dependence of the
bispectrum, the maximum likelihood values for the bias parameters 
are very similar to those in Table~1, 
but with larger error bars. As we increase the number of
eigenmodes from 2 to 203, the likelihood contours slowly contract
(these additional eigenmodes all have signal to noise less than 
unity in our fiducial model); the maximum likelihood bias values for
the 2~Jy catalog hardly change, whereas in the 1.2~Jy survey the
maximum likelihood slowly oscillates in the linear bias direction
within the range $1/b_1=1.0$--1.3.

Since these additional eigenmodes carry information on the scale 
dependence of the reduced bispectrum, they can help to constrain 
non-Gaussian initial conditions, as exemplified by the 
$\chi^2$ model \citep{LM97,AMM97,P97,P99a,P99b}, in which the linear
density field is the square of a Gaussian field. As a result, the
initial bispectrum is non-zero and has a different scaling, 
$B_i \propto P_i^{3/2}$, and shape dependence from the gravitionally
induced bispectrum in \eq{B123}.  However, in order to compare with
observations, linear extrapolation of the initial conditions is
insufficient, except at extremely large scales or high redshift, which
are not probed by the \IRAS surveys. Taking into account gravitational
corrections, one finds that the overall scale-independence of $Q$ is
partially restored by gravity over the scales we probe; however, the
amplitude of $Q$ is roughly two times larger than in the Gaussian
case.
The configuration dependence resembles that of the Gaussian case, 
but with residual dependences on the ratios $k_i/k_j$, which would
show up as a rather strong dependence on the variable $r$ in \eq{Qvar}
\citep{S00a}. In order to test this model against observations, we
must take into account the effects of biasing as well, which are
non-trivial for general non-Gaussian models.  For the particular case
of $\chi^2$ initial conditions, however, a simple result follows,
\beq
Q_g = \frac{1}{b_1} Q + \frac{b_2^{\rm eff}}{b_1^2},
\label{Qbiaschi}
\eeq 
where $b_2^{\rm eff} \approx (3-8/\sqrt{3}\pi) b_2 \approx 1.53 \,
b_2$ \citep{S00a}. As we did in the Gaussian case, we shall assume
that the same relation approximately holds in redshift-space as
well. We have run 2LPT realizations and created mock 1.2~Jy catalogs
for $\chi^2$ initial conditions, as described in Paper~I. We apply the
same likelihood analysis as in the Gaussian case, with appropriate
change for the mean, variance, and PDF of the bispectra in
\eq{like}. The results of this analysis are shown in
Figure~\ref{likeNG}. If we consider only the two highest $S/N$
eigenmodes, which probe the overall amplitude and configuration
dependence of the bispectrum, we obtain the $68\%$ contour shown in
the bottom panel, which gives $1/b_1=1.42^{+0.34}_{-0.49}$ and
$b_2/b_1^2=-1.40 \pm 0.58$, consistent with the fact that the overall
amplitude of $Q$ is larger than in the Gaussian case, and the
configuration dependence is about the same. However, including the
remaining eigenmodes, which are sensitive to $r$ and $k$ dependence,
the maximum likelihood is driven to $1/b_1=0$, in an attempt to cancel
the $r$ and $k$ dependence in the $\chi^2$ model not seen in the
data. If we consider that \IRAS galaxies are not biased at large
scales ($1/b_1 \geq 1$), then the 1.2~Jy bispectrum results rule out
$\chi^2$ initial conditions at the 95\% confidence level. We note that
this model may be considered `strongly' non-Gaussian; models with
substantially `weaker' initial non-Gaussianity (e.g., with negligibly
small initial three- and four-point correlations) are not excluded by
this analysis.

We can also use this data to constrain the Lagrangian local bias model.
This model also introduces additional dependence of the bispectrum 
on $r$, through the last term in \eq{Qbiaslag}, although with a much
smaller amplitude than in the $\chi^2$ model. We can therefore try to
test whether the local Lagrangian or the Eulerian model is preferred. 
We have repeated the likelihood
analysis using \eq{Qbiaslag} in place of \eq{Qbias} to incorporate the
change in the mean of $Q$ as a function of biasing parameters. We find
that the goodness of fit, given by the value of the likelihood function
at its maximum, does not change from one model to the other by more than
50\%. The largest change is obtained for the 2~Jy survey, shown in
Figure~\ref{like_LAG} , where we compare the Eulerian result (same as in
Figure~\ref{likeG}) to the Lagrangian case (for the $\Lambda$CDM parameters). 
A similar behavior as in the $\chi^2$ model is observed: the
contours shift towards higher bias, as listed in Table~1.
 These values of the bias parameters are consistent with those 
obtained in the Eulerian model. A larger dataset is 
required to distinguish between these two biasing schemes.

\renewcommand{\thefootnote}{\fnsymbol{footnote}}

\section{Conclusions}

We have presented results for the reduced bispectrum $Q$ of three
\IRAS redshift catalogs, computed using the optimal weighting scheme
of FKP, after subtracting contributions to the power from discreteness
and from the shape of the selection function.
That $Q$ is nonzero, positive, and generally of the shape predicted by
gravitational instability is unmistakable. The ability to extract
values for bias parameters is marginal; however, after taking into
account non-linear redshift distortions and finite volume effects,
likelihood analysis yields reasonable values for the parameters of a
local Eulerian bias model, $1/b_1 =1.32^{+0.36}_{-0.58}$,
$1.15^{+0.39}_{-0.39}$ and $b_2/b_1^2=-0.57^{+0.45}_{-0.30}$,
$-0.50^{+0.31}_{-0.51}$, for 2~Jy and 1.2~Jy respectively.  This
implies that \IRAS galaxies trace dark matter with decreasing
effective bias as a function of the mass density fluctuation
amplitude, consistent with their being under-represented in the cores
of clusters.
We have also considered constraints on non-Gaussian initial
conditions, using as an example the $\chi^2$ model, in which the
bispectrum develops a different amplitude and scale dependence than in
the Gaussian case.  If \IRAS galaxies are {\it not} positively biased
at large scales (i.e., if $b_1^{IRAS} \leq 1$), then $\chi^2$ initial
conditions are ruled out at the 95\% confidence level.

The likelihood contours show that one combination 
of the bias parameters is known to better accuracy: 
the long axis of the likelihood contours in Figures 
\ref{likeG}--\ref{like_LAG} follows tracks of roughly constant 
skewness, $ S_3 = 3.25/b_1 + 3 b_2/b^2 $ for $ n = -1.4 $.
Our values inferred for $S_3$ are consistent with previous work 
on the skewness of \IRAS galaxies.
For the 1.2~Jy sample, \citet{BSDFYH93} obtained 
$ S_3 = 1.5 \pm 0.5 $; for the same catalog, \citet{KS98},
who include a finite volume correction but use a different estimator, 
obtained $ S_3 = 2.83 \pm 0.09 $.
Our results translate via \eq{S3} (assuming $n=-1.4$) 
into a finite-volume corrected value 
$S_3 \approx 2.25 \pm 0.9$, comparable to that of \citet{KS98}.  
To estimate the error bar in the skewness we assume that $S_3$ 
corresponds to sum of $Q$ over all triangles.
The error is then roughly given by three times
the width in the best determined direction in the ellipse in
Figure~\ref{likeG}, $\pm 0.3$.
An approximate estimate of $S_3$ that does not include the finite volume
corrections can be obtained directly from Figure~\ref{likeG}, 
by using the fact that at the $68\%$ limit, $Q/b_1+b_2/b_1^2$ 
is in the range 0.2--0.8; this implies
 $S_3 \approx 3 Q_g \approx 1.5 \pm 0.9$, consistent with
\citet{BSDFYH93}. These are only rough estimates, as they ignore the
skewness of the bispectrum (and $S_3$) PDF.

Our results have interesting implications for the relative
clustering of \IRAS and optically selected galaxies. From the ratio of
power spectrum amplitudes between the Stromlo-APM survey and the
combined QDOT-1.2Jy survey, \citet{TE96} conclude that $
b_1^{opt}/b_1^{IRAS} \approx 1.2 $, while from the analysis of the APM
projected three-point function, \citet{FG99} conclude that $b_1^{opt}
\approx 1$.  These results are consistent with our findings for
$b_1^{IRAS}$.  Future galaxy surveys will be able to quantify this
issue with a greater level of accuracy.

\acknowledgements

We thank Stephane Colombi, Enrique Gazta\~{n}aga, Lam Hui, Simon Prunet, 
and Matias Zaldarriaga for useful discussions.  Research supported in
part by the NSF EPSCoR Grant at Kansas and the University of Kansas
GRF, by the DOE and NASA grant NAG 5-7092 at Fermilab, and by 
NASA NAG5-2835 at the University of Florida.
Parts of this work took place at the Aspen Center for Physics.

\clearpage

\begin{deluxetable}{rccc}
\tablewidth{0pt}
\tablecaption{Maximum Likelihood Estimates of Bias Parameters 
\label{tbias}}
\tablehead{\colhead{Sample} & \colhead{Model} & %
\colhead{$1/b$} & \colhead{$b_2/b^2$} }
\startdata
2~Jy   & $\Omega=0.3$ & $ 1.32^{+0.36}_{-0.58} $ & $-0.57^{+0.45}_{-0.30} $ \\
1.2~Jy & $\Omega=0.3$ & $ 1.15^{+0.39}_{-0.39} $ & $-0.50^{+0.31}_{-0.51} $ \\
2~Jy   & $\Omega= 1 $ & $ 1.16^{+0.38}_{-0.46} $ & $-0.52^{+0.40}_{-0.20} $ \\
1.2~Jy & $\Omega= 1 $ & $ 1.05^{+0.30}_{-0.48} $ & $-0.44^{+0.40}_{-0.33} $ \\
2~Jy & Lag. bias ($\Omega=0.3$)& $ 1.08^{+0.08}_{-0.54} $ & $-0.48^{+0.72}_{-0.03} $ \\
\enddata
\end{deluxetable}

\clearpage

 \plotone{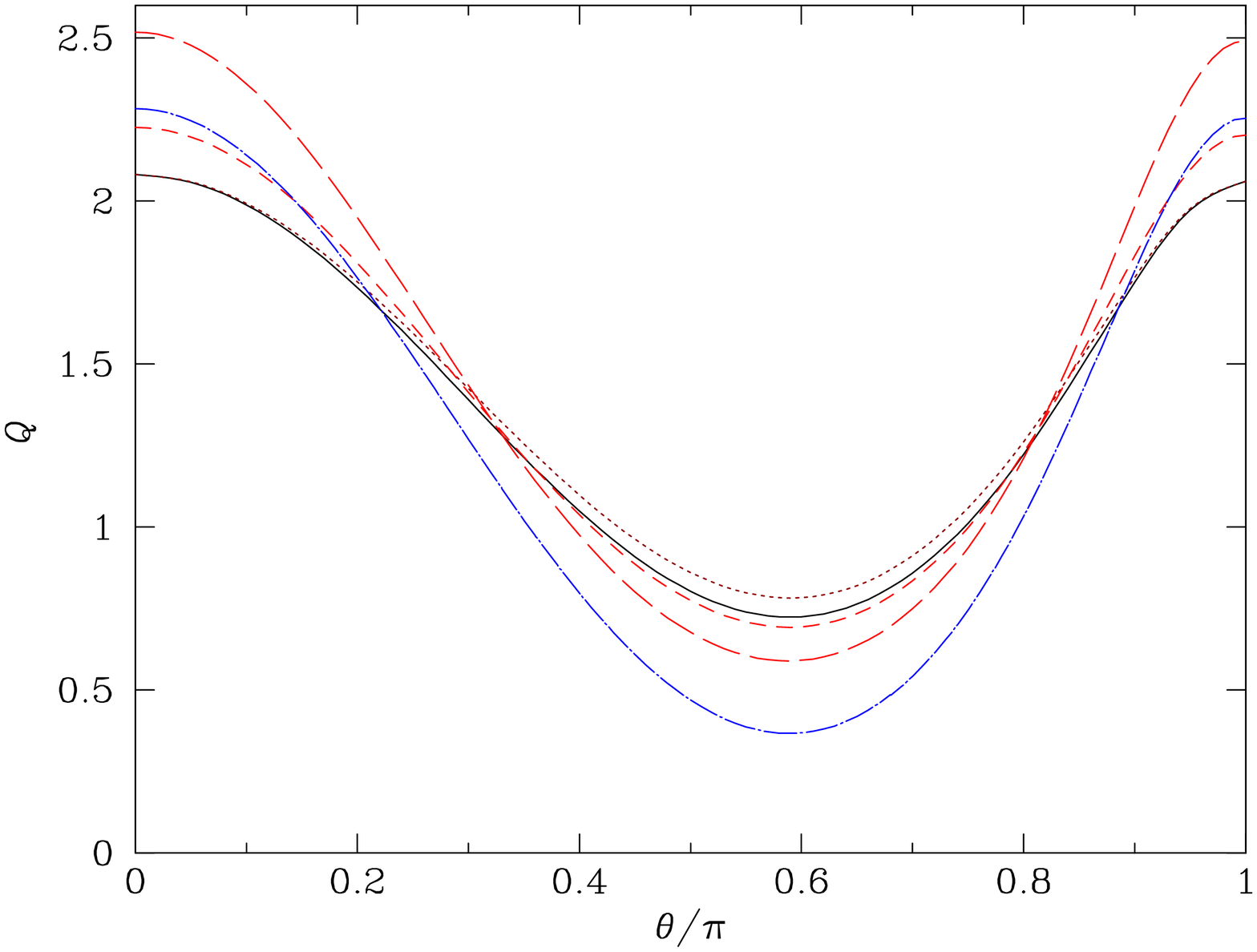} \figcaption {The dependence of $Q$ on configuration
shape expected from perturbation theory, for configurations with 
$ k_2/k_1 = {1 \over 2} $ separated by angle $ \theta $, for a
power spectrum $ P(k) \sim k^n $ with $ n = -1.4 $.  The solid curve
shows the result in real space, the short-dashed curve in redshift
space for $ \Omega_m = 0.3 $, and the long-dashed curve in redshift
space for $ \Omega_m = 1 $.  For $ \Omega_m \to 0 $ the dotted curve
applies in both real and redshift space.  The dot-dashed curve shows
the effect of a local bias as in equation (\protect{\ref{Qbias}}),
with $ 1/b_1 = 1.25 $ and $ b_2/b_1^2 = -0.5 $.
\label{Qpt}}

 \plotone{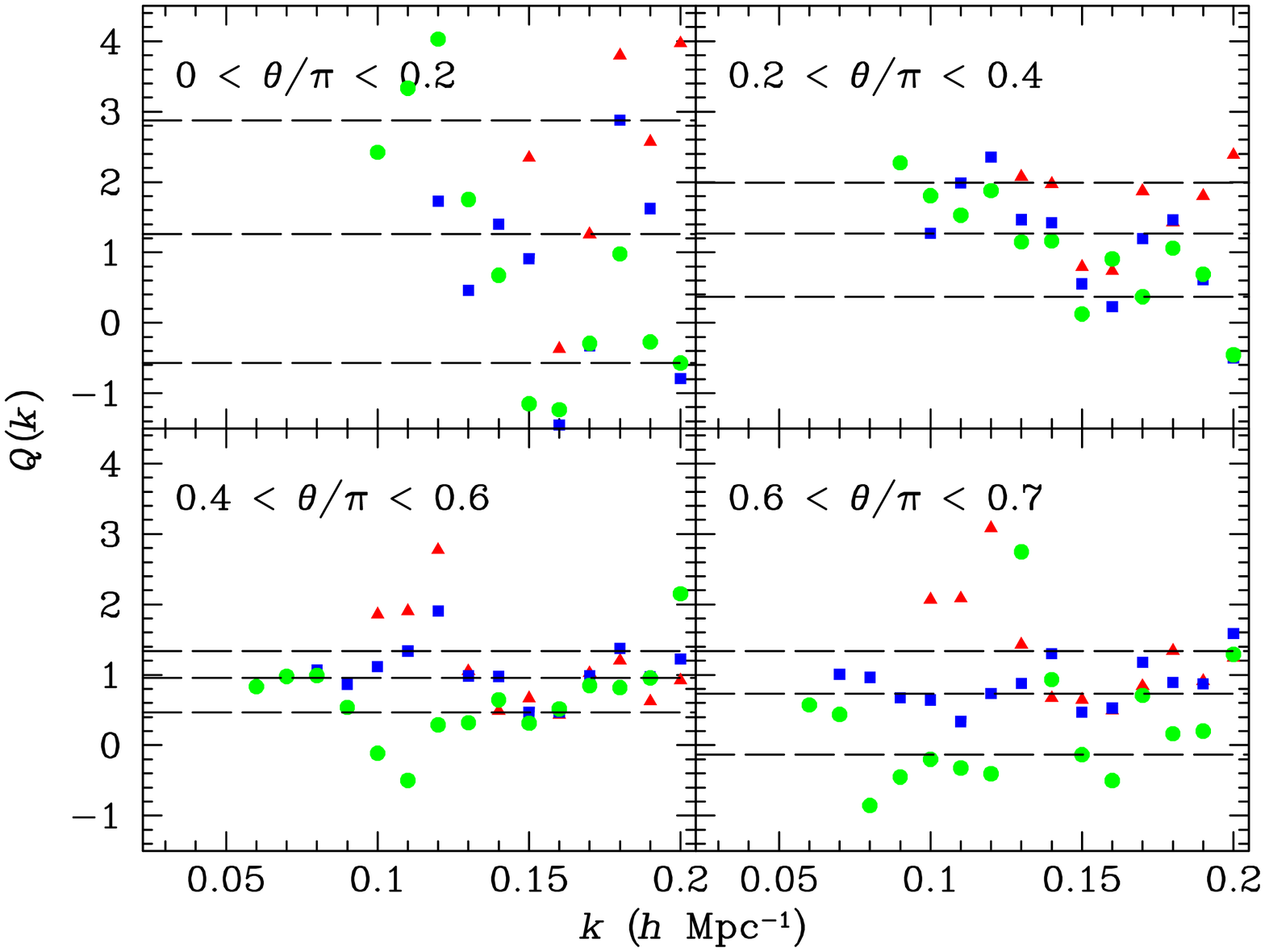} \figcaption {$Q$ vs.~$k = k_1$ for the QDOT
catalog for triangles with two sides of a specified ratio 
separated by angle $ \theta$.  
The four panels show bands in $ \theta $, from nearly colinear, 
$ 0 \leq \theta_{23}/\pi < 0.2 $ (upper left), 
through $ 0.6 \leq \theta_{23}/\pi \leq 0.7 $ (lower right).
In each panel, 
Triangles show $ k_2/k_1 = 0.25 $--0.5; 
squares show $ k_2/k_1 = 0.5 $--0.75; 
circles show $ k_2/k_1 = 0.75 $--1.0.
The dashed lines in each panel show the median and
68\% range of the points.
\label{QkQDOT}}

 \plotone{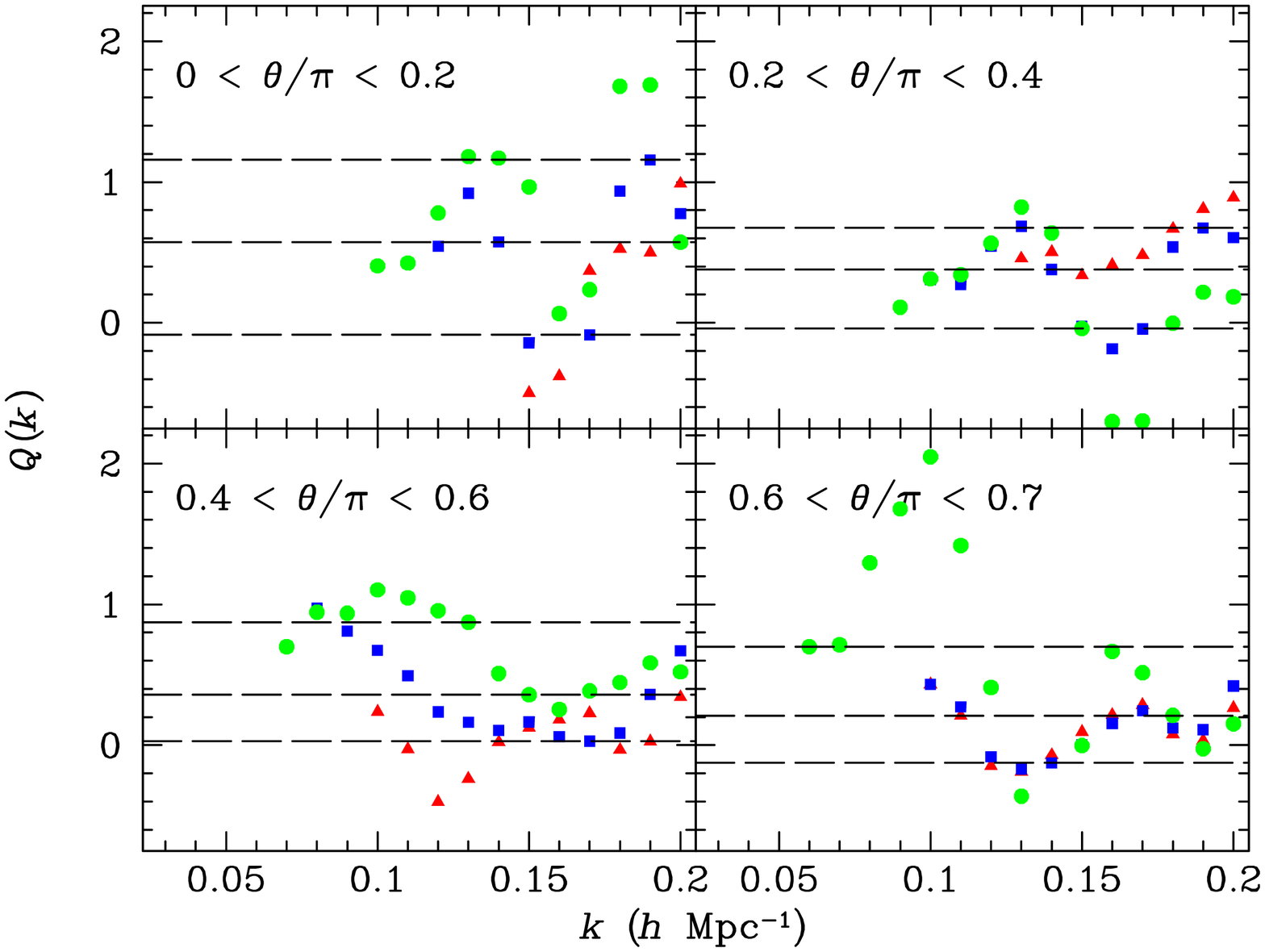}
\figcaption {Same as in Fig.~\protect{\ref{QkQDOT}}, 
for the 2~Jy catalog.
\label{Qk2Jy}}

 \plotone{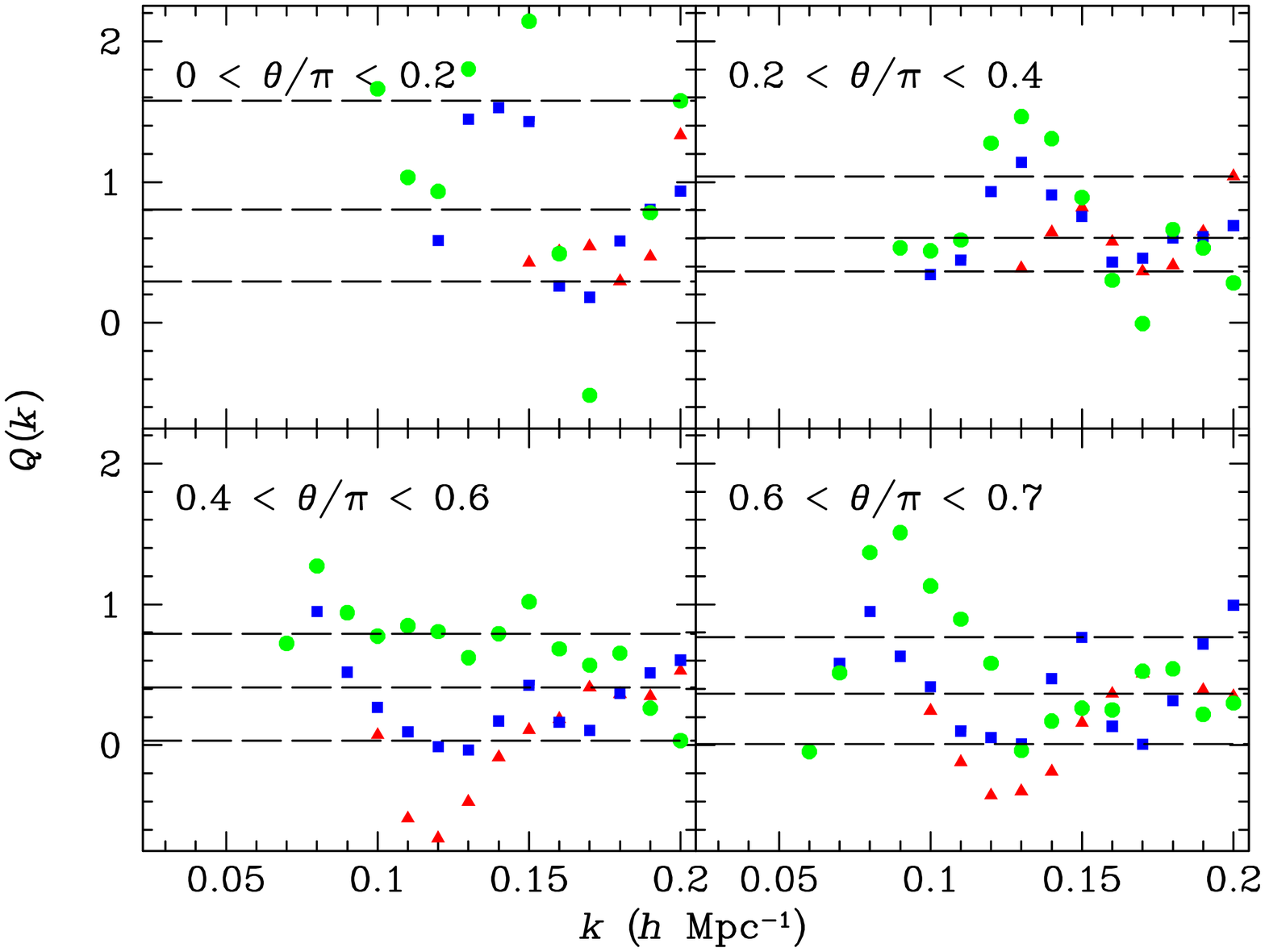} 
\figcaption {Same as in Fig.~\protect{\ref{QkQDOT}}, 
for the 1.2~Jy catalog.
\label{Qk1.2Jy}}

 \plotone{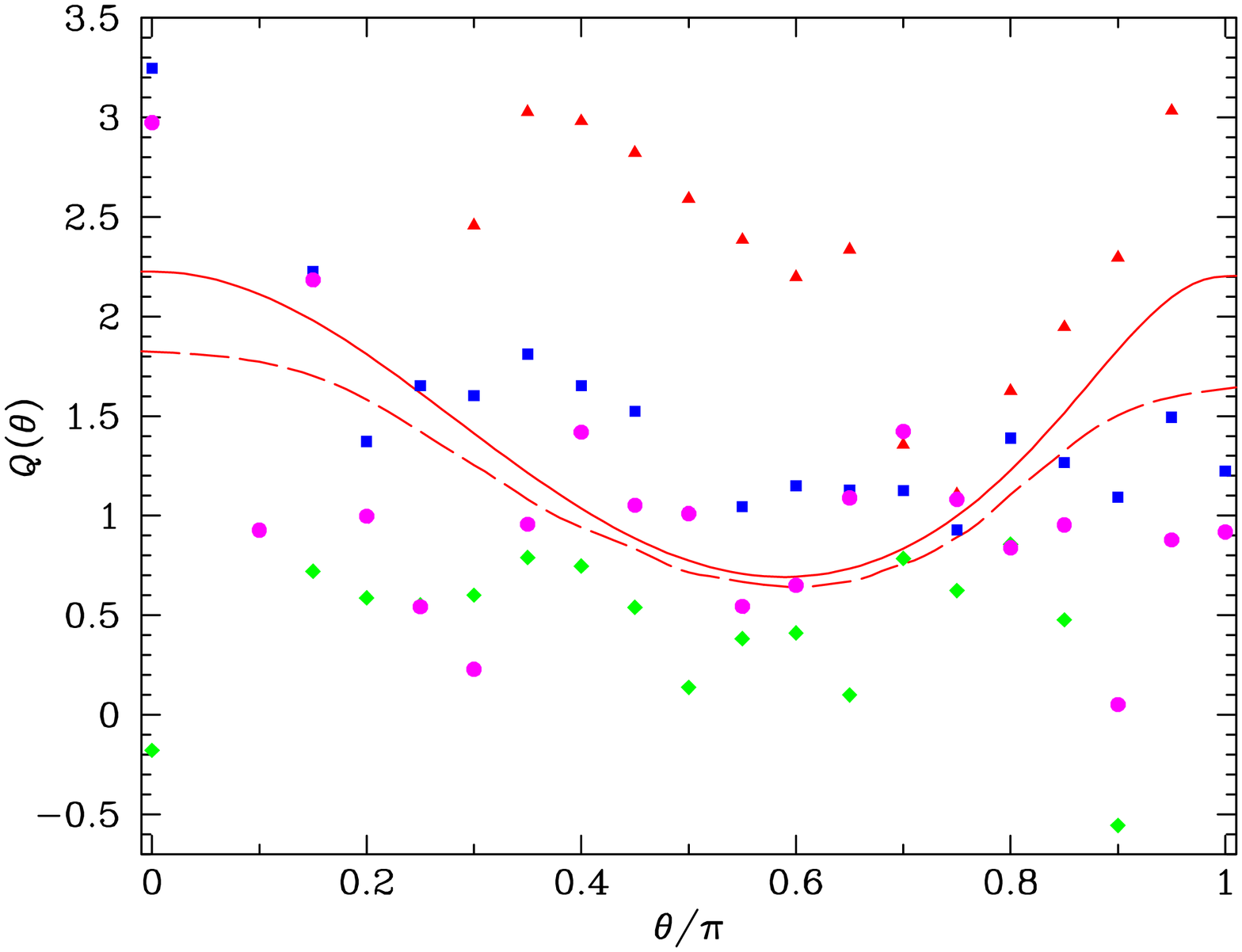} 
\figcaption{$Q$ vs. $\theta $ for the QDOT catalog for 
triangles with $ 0.05 \leq k \leq 0.2 \kMpc $ and with two sides of
ratio 0.4--0.6 separated by angle $ \theta $.
The solid curve shows $Q$ expected from perturbation theory for 
spectral index $ n = -1.4 $ in equation (\protect{\ref{Qred}}) 
for  $ \Omega = 0.3 $.  
The long-dashed curve shows $Q$ in redshift space averaging many 
2LPT synthetic realizations using the full LCDM power spectrum, 
showing the nonperturbative corrections discussed in \S 4.1.
Symbols show results for bands in $k_1$: triangles for 
$ k_1 = 0.1 $--$ 0.125 \kMpc $; squares, 0.125--$ 0.15 \kMpc $; 
diamonds 0.15--$ 0.175 \kMpc $; and circles 0.175--$ 0.2 \kMpc $.
\label{QaQDOT}}

 \plotone{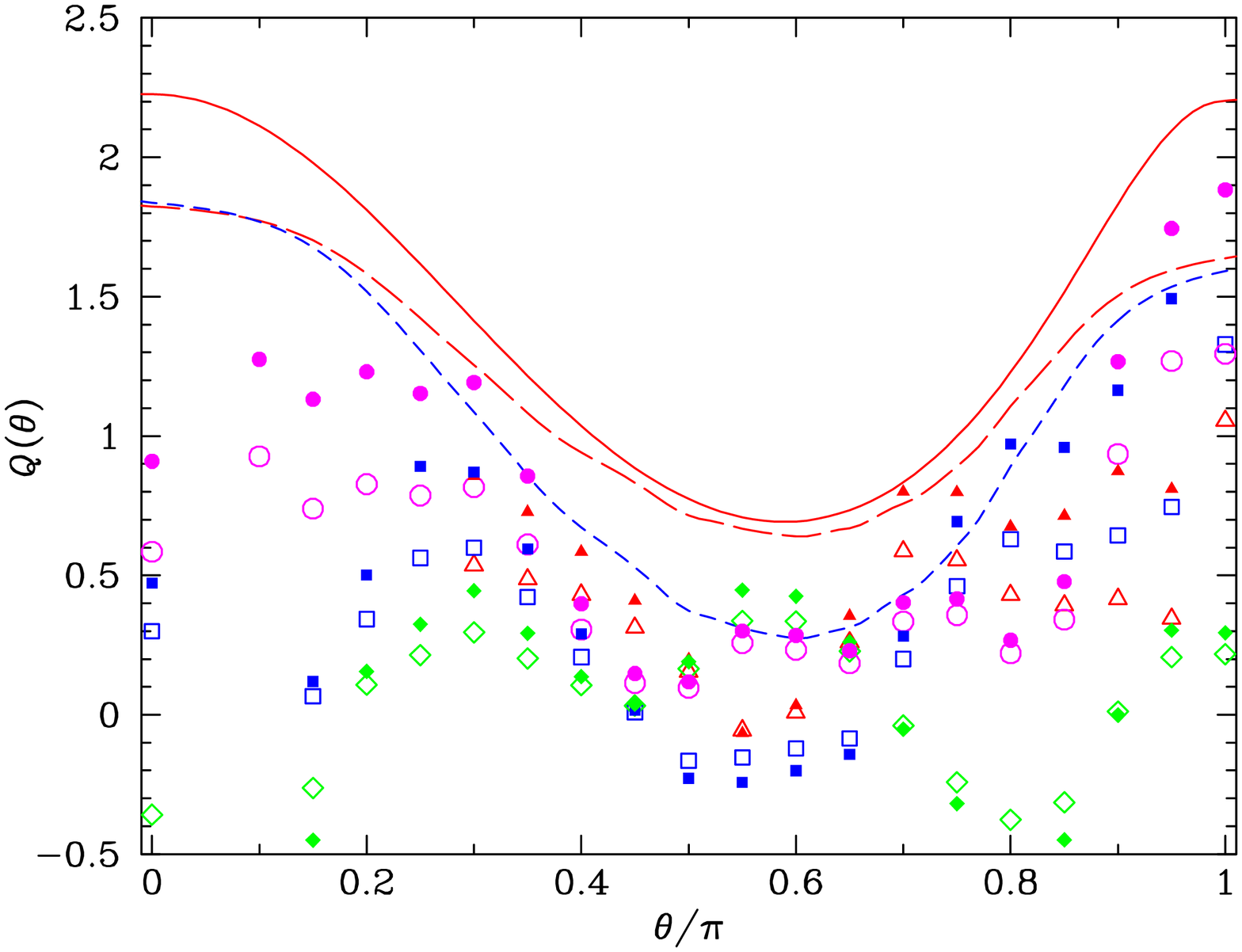} \figcaption{Same as Fig.~\protect{\ref{QaQDOT}}, 
for the 2~Jy catalog.  Open symbols show raw results; filled symbols
after correcting for the average finite volume bias.
The short-dashed line shows the result expected for 
$ 1/b = 1.32 $, $ b_2/b^2 = -0.57 $ applied to the 2LPT result.
Note that the short-dashed line is {\em not} obtained by fitting
directly to the filled symbols; since the PDF of $Q$ is positively
skewed low values of $Q$ are more probable than high values. 
\label{Qa2Jy}}

 \plotone{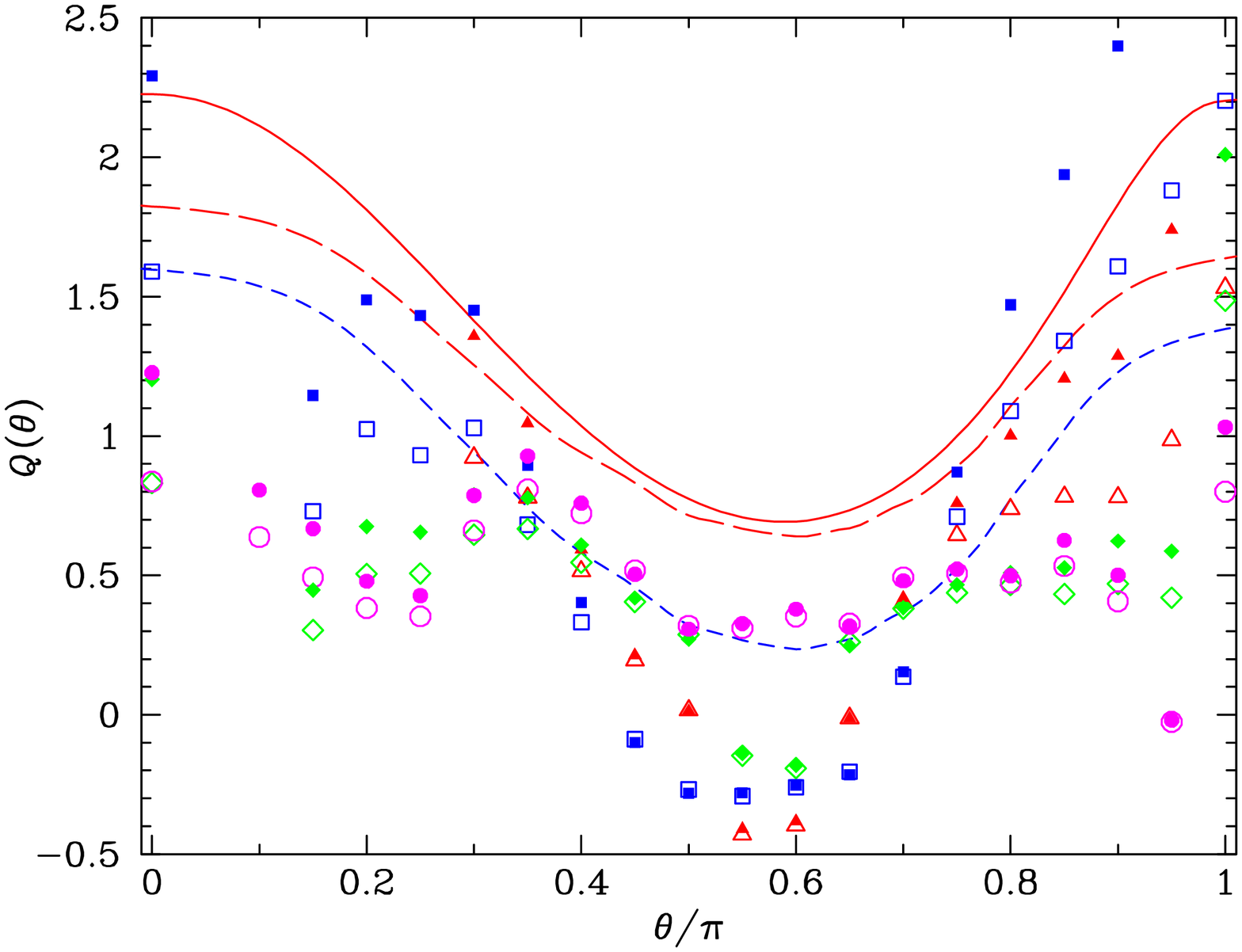}
\figcaption{Same as Fig.~\protect{\ref{Qa2Jy}}, 
for the 1.2~Jy catalog.
The short-dashed line shows the result expected for 
$ 1/b = 1.15 $, $ b_2/b^2 = -0.50 $ applied to the 2LPT result.
\label{Qa1.2Jy}}

 \plotone{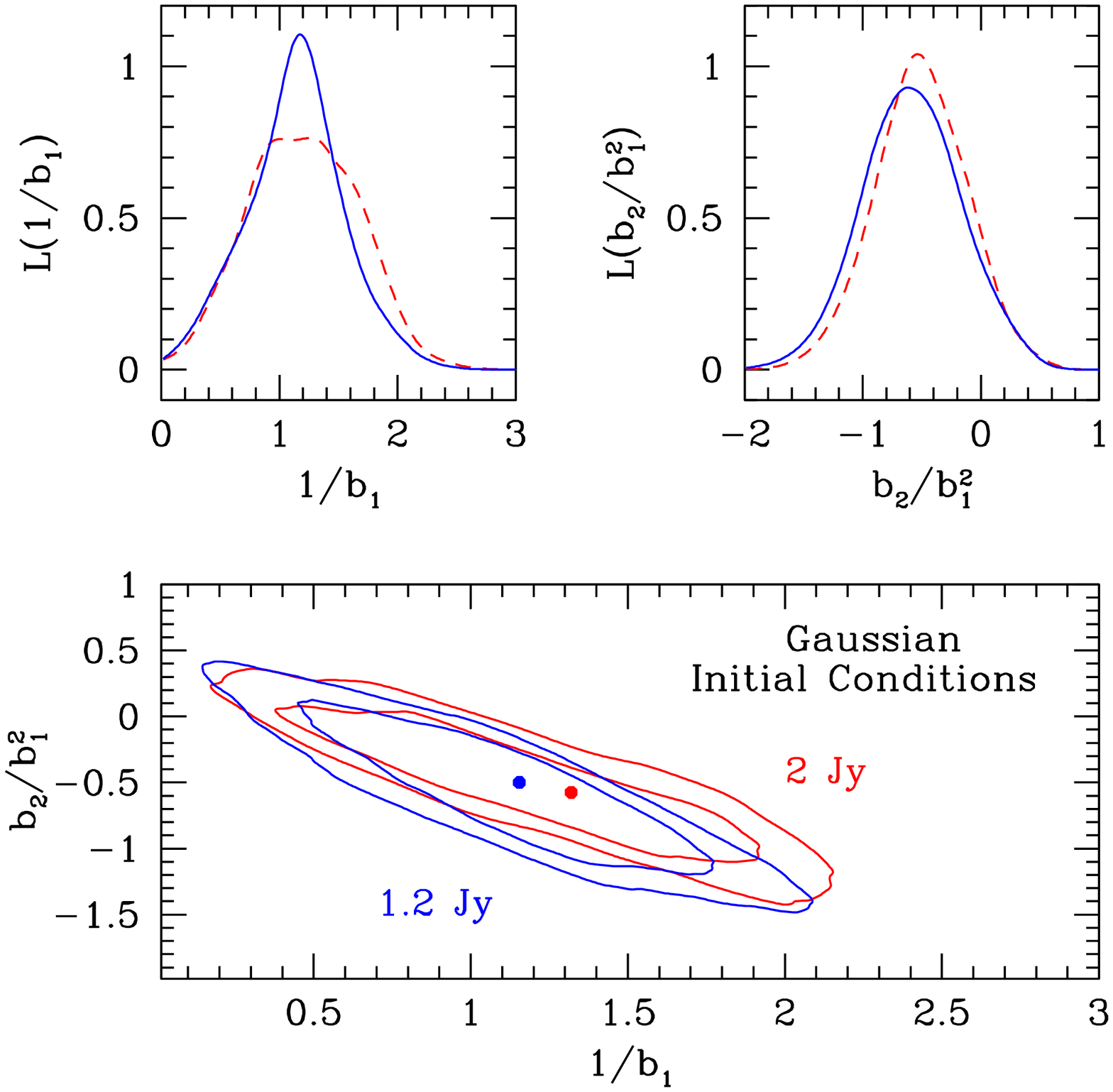} 
\figcaption{
The bottom panel shows $68\%$ and
$90\%$ likelihood contours for bias parameters assuming Gaussian
initial conditions and $\Omega_m=0.3$ 
for the 2~Jy and 1.2~Jy surveys. In the upper
panels we show the marginalized likelihoods (normalization arbitrary)
for 2~Jy (dashed) and 1.2~Jy (solid) catalogs. 
\label{likeG}}

 \plotone{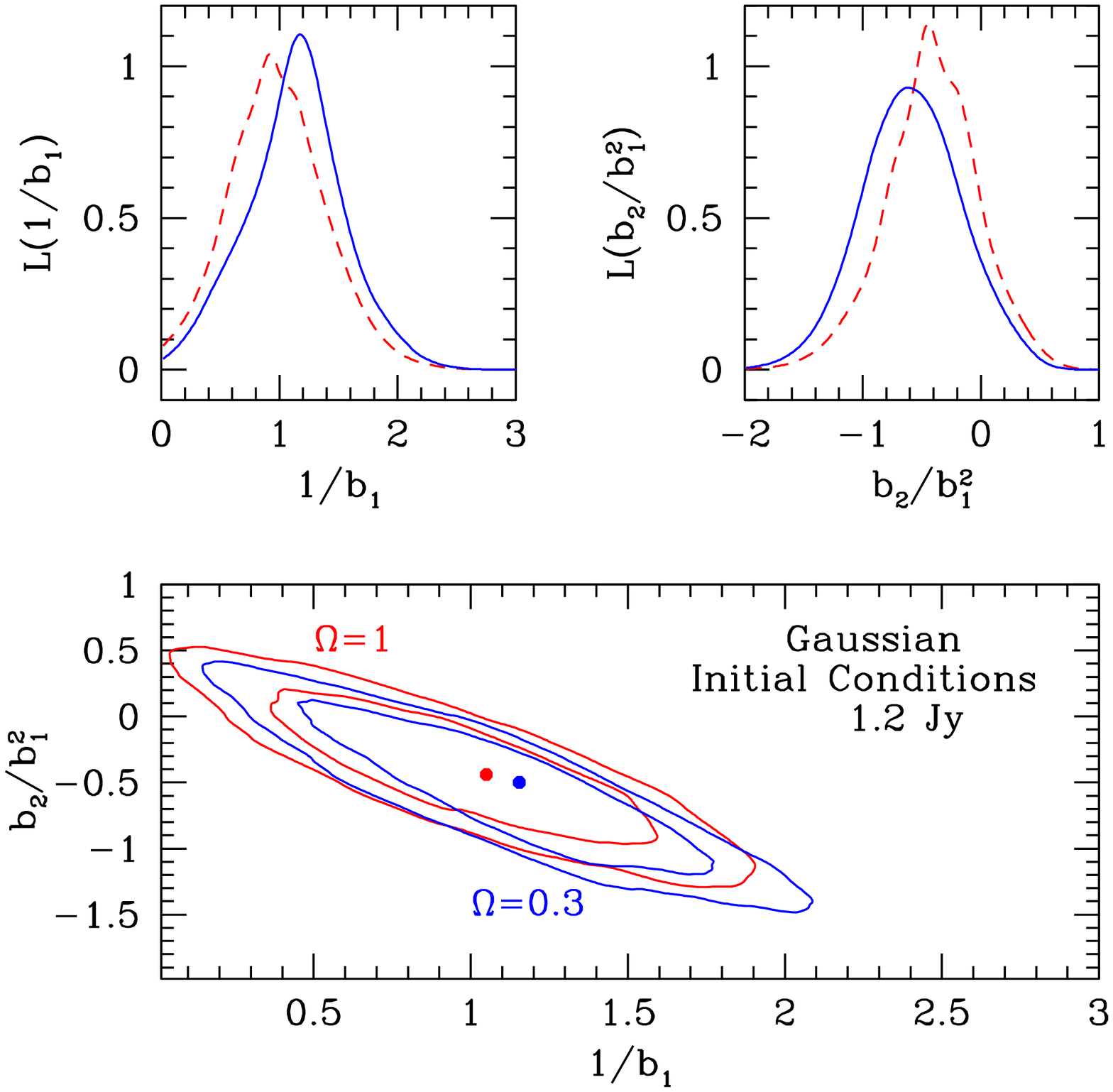} 
\figcaption{
Same as Fig.~\protect{\ref{likeG}}, but showing the dependence of the 
results on $\Omega_m$  
in the 1.2~Jy catalog. In the upper panels, we show the marginalized
likelihoods for $\Omega_m=1$ (dashed) and $\Omega_m=0.3$ (solid).
\label{likeGo}}

 \plotone{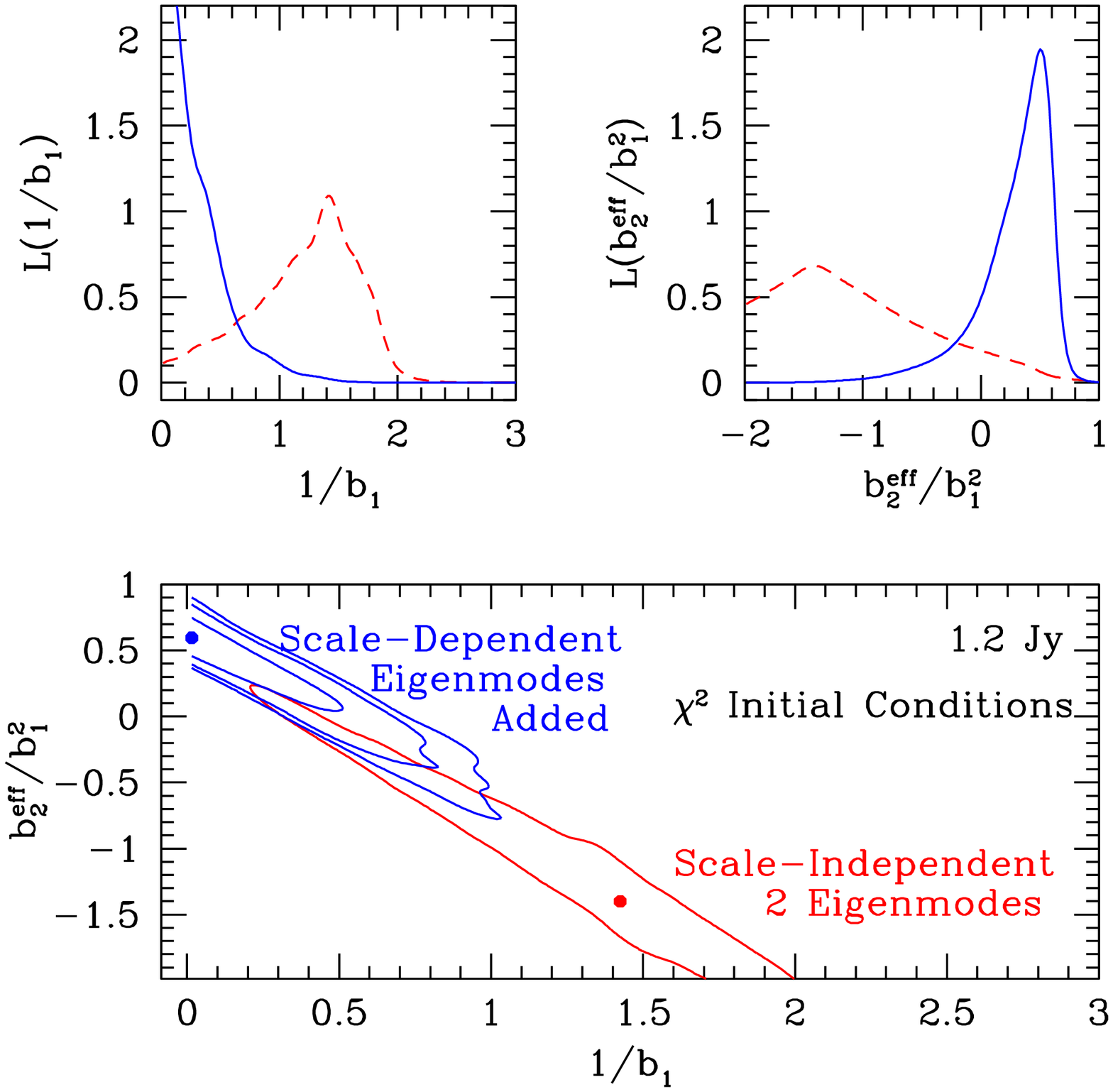}
\figcaption{
Likelihood analysis for $\chi^2$ initial conditions in the 1.2~Jy
survey. Contours for the 2 highest $S/N$ eigenmodes are $68\%$; contours for 
all eigenmodes are $68\%$, $90\%$, and $95\%$. Upper panels show
marginalized likelihoods for 2 highest (dashed) and all (solid)
eigenmodes. 
\label{likeNG}}

 \plotone{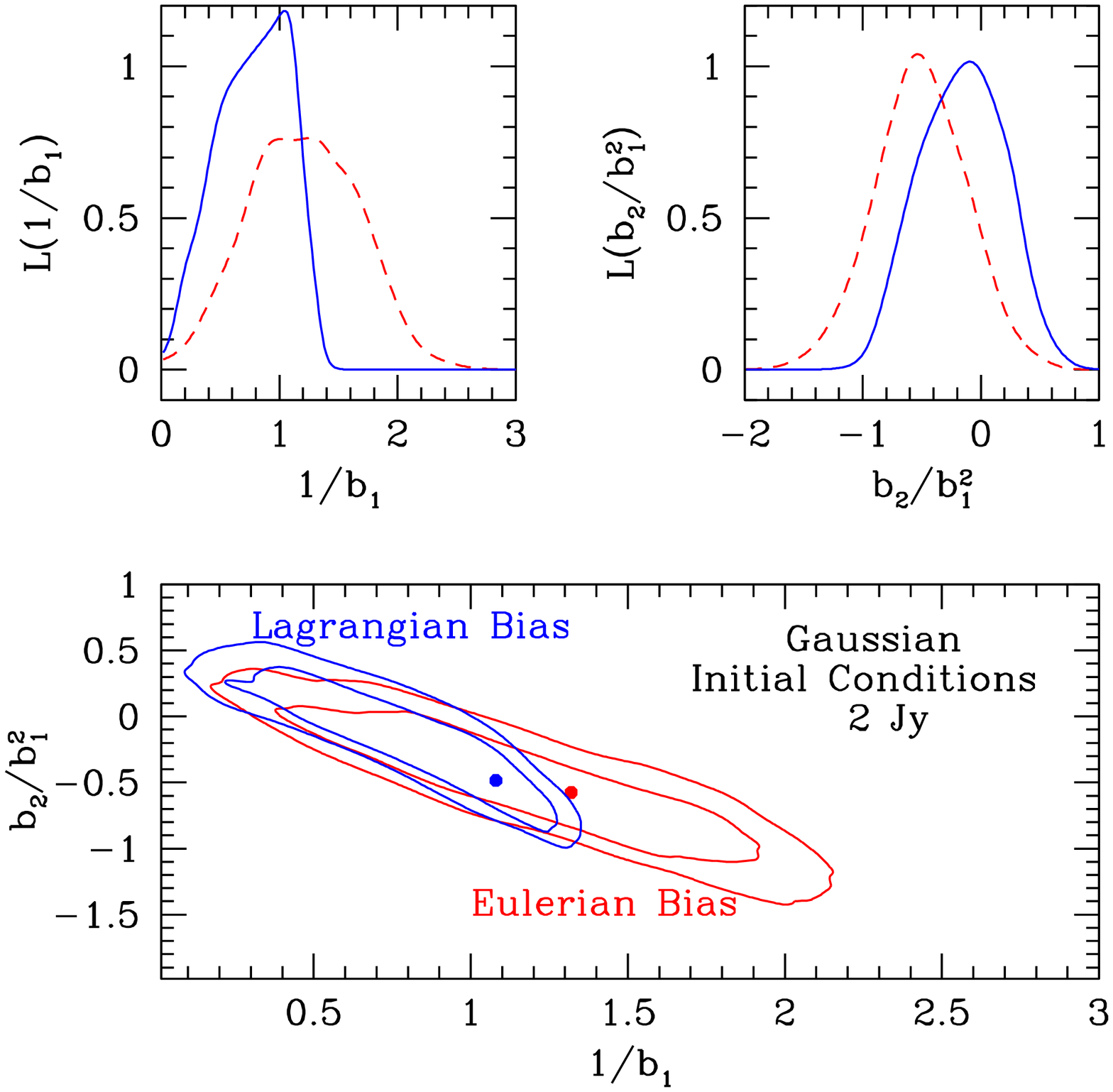}
\figcaption{
Likelihood analysis for the Lagrangian local bias model in the 2~Jy
survey.
Upper panels show marginalized likelihoods for Eulerian (dashed) 
and Lagrangian (solid) bias. 
\label{like_LAG}}

\end{document}